# CHAPTER TWO

# SOFTWARE DEVELOPMENT INDUSTRY IN EAST AFRICA: KNOWLEDGE MANAGEMENT PERSPECTIVE AND VALUE PROPOSITION

**Karanja Evanson Mwangi, Lawrence Xavier Thuku and John Patrick Kangethe**

karanjae@gmail.com; lthuku@symphony.co.ke; karanjae@gmail.com

**Abstract.**

*Increased usage of internet has contributed immensely to the growth of software development practice in East Africa. This paper investigates the existence of formal KM (Knowledge Management) initiatives in the Software industry such as creation of virtual communities ( Communities of practice and communities of interest); expert localization and establishment of knowledge taxonomies in these communities; the knowledge transfer and sharing processes; incubation and Mentorship; collaborative software development and their role in creating entreprenuership intiatives and providing a building block towards the knowledge economies. We propose a hybrid framework for use in KM intiative focusing on Software Development in East Africa.*

### Introduction

Researchers in the field of knowledge management acknowledges the complexity of isolating and defining knowledge, its constituents and their dependencies (Bergeron, 2003), (Bouthillier & Shearer, 2002) . Due to the interdisciplinary nature of Knowledge Management (KM), the attempts to define these various concepts and constituents slightly differ depending on the discipline of influence and context.  For the purposes of this paper, we identified the following constituents: Data, information, knowledge and instrumental understanding.  A review on KM literature shows that there exists variances and overlaps in the definitions of these constituent across the authors.(Meadow, Boyce, & Kraft, 2000) defines data as strings of elementary symbols, such as digits and letters. As they argue that information is generally made up of evaluated or useful data. Knowledge has higher degree of validity and has "characteristics of information shared and agreed by a community". Meadow, et al. relates instrumental understanding to intelligence which they define as a measure of reasoning capacity.(Wiig, 1999) defines information as organized data and knowledge as a set truth and belief. (Bergeron, 2003) argues that data are numerical quantities drawn from observation, experiment or calculation whereas Information is applied data: - "collection of data and associated explanations, interpretations, and other textual material concerning a particular object, event, or process". Bergeron introduces metadata as a link between information and knowledge, which he defines as: "information about the context in which information is used".  Knowledge is illustrated as a mix of metadata and awareness of the context which metadata can successfully be applied.

(Zack, 1999) defines data as observations and fact out of a context that has no direct meaning and information is data within a meaningful context. (Liew, 2007), (Govil, 2007) observed that data must be processed (to be put into meaningful context) to obtain information that a decision can be based on. Knowledge is derived from validated information and differentiated through experience whereas instrumental understanding is the utilization of accumulated knowledge. Therefore, there is need for proper data management, information management and knowledge management in their hierarchical relationship so as to realize the aim of KM (Govil, 2007; Hick, 2006).

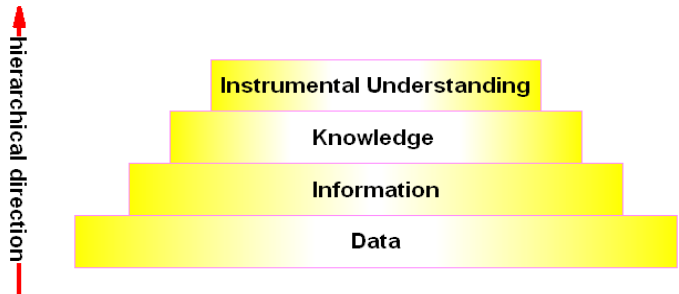

Figure 1: Knowledge Continuum

In general the above definitions highlight the overlap between the constituents of knowledge. There seems to be a consensus among the authors on three issues namely; the relationship between adjacent constituents, the validity of constituents is dependent on the context and the hierarchical direction (Data at the bottom of the hierarchy and instrumental understanding at the top) as illustrated in **Figure 1 above (Knowledge continuum).**

**Background Study of Knowledge Management**
*The Multi- Facet nature of Knowledge Management*
The practice of KM has been there for a long time mainly in the informal way (Pandya & Xiaoming, 2003). There are numerous working definitions of KM (Bouthillier & Shearer, 2002) cited the work of (Hlupic, Pouloudi, & Rzevski, 2002) which identified 18 definitions of KM in different contexts . The working definition used by the authors of this paper is:*"Is the ability of a community to create, validate, audit, share knowledge using appropriate technologies to gain competitive advantage"*

The KM thinking and praxis is informed and influenced by various disciplines. (Kakabadse, Kakabadse, & Kouzmin, 2003) study infers as follows on the nature and sources of influence:*" philosophy, in defining knowledge; cognitive science (in understanding knowledge workers); social science (understanding motivation, people, interactions, culture, environment); management science (optimizing operations and integrating them within the enterprise); information science (building knowledge-related capabilities); knowledge engineering (eliciting and codifying knowledge); artificial intelligence (automating routine and knowledge-intensive work) and economics (determining priorities)."*

Knowledge management is a multi- facet discipline stretching across numerous economic sectors. Organizations within those sectors have differing approaches based on theoretical perceptions or practical experiences on how knowledge can be effectively managed i.e created, validated, transferred and re-used.

*Km And Intellectual Capital*

As organization in these sectors attempt to move the knowledge realms from cognitions and abilities of individuals to a vital transitional asset , they are faced with the challenge of organizing and leveraging its intellectual capital faster than their competition (Bontis, 2001) . Different studies on intangible assets identify three major components of intellectual capital namely; human capital, customer capital, and structural capital. (Bergeron, 2003), (José, 2003) , (Edvinsson & Malone, 1997), (Sullivan, 2000), (Sveiby, 1997), (Kaplan & Norton, 1996).

Pike, Rylander, & Roos,( 2002) uses the term relational capital to refer to customer capital. VanBuren, (1999), Hsu & Mykytyn, (2006) isolates the structural capital to innovation capital and process capital and argues that there exist an intuitive link between the various components of intellectual capital. The effective management of these intellectual capital components and their inter-relationship is an important step towards organizational learning and market leadership.

Numerous researchers have investigated the KM and Intellectual capital issues from an organizational perspective. In this paper we extend the concept of intellectual capital and KM beyond the organization view to a community view of knowledge that spans among different organizations and individuals who practice or have interest in software development. Recent advances in software development especially emergence of active communities (localized or virtual) have necessitated critical consideration of KM as an integral part of the practice and success of software industry (Hemetsberger & Reinhardt, 2003).

*Knowledge Management practice in East Africa*

The East Africa Region is made up of three nations; Kenya, Uganda and Tanzania. KM has been going on informally and intuitively in the East African region. Empirical study based on a Kenyan perspective that shows of KM initiatives are firm based (Mosoti & Masheka, 2010). Organizations in the East Africa region uses in-house approaches or strategic partnerships as ways of realizing Knowledge management, however to measure the effectiveness of these practices is difficult due existing organizational culture and vocational reinforcers that induce the notion that knowledge sharing among organizations in the similar or complementary industries may reduce their competitive advantage and market leverage. Formal KM is an emergent area with great value proposition in Africa (Karanja, 2010) . Knowledge Management Africa (KMA) is one of the new initiatives that aim at driving KM initiatives in Africa.

The African Medical and Research Foundation (AMREF) is an organization headquartered in Nairobi – Kenya with operations in seven African countries i.e. Kenya, Uganda, Ethiopia, Somalia, Tanzania, South Sudan, and South Africa. AMREF is facilitating a community participatory approach to knowledge Management in the health sector. AM-REF has partnered with local communities, health system formulators and governments with an aim of realizing right to health for all (Ireri & Wairagu, 2007).

Kora, (2006) evaluates the viability of Information and Communication Technologies (ICT) as a KM strategy in rural development in Tanzania. None of the KM research initiatives in East Africa has formally focused on Software development, despite being a prominent contributor in the region's emerging knowledge economy.

*Knowledge Management Models*
Kakabadse et al., (2003) while extending the work of Swan & Newell, (2000) provided for Five useful models of KM , where each model treat KM initiatives differently. They identified the models as follows:

Philosophy based model – it's concerned with the epistemology of knowledge or what constitutes knowledge, the relationship of the constituents and other notions such as truth, justification, causation, doubt and revocability. The model provides a high level perspective that requires reflections in areas of practice. It's mainly grounded on Socratic view of knowledge as justified true belief and wisdom as highest constituent in the knowledge continuum. Proponents of this model argue that KM needs not be technology centred.

Cognitive Model –: this model is rooted on recognition of knowledge as an economic asset. It focuses on organisational perspective of knowledge and considers ICT as an enabler of the knowledge management process. (Swan & Newell, 2000), (Zack, 1999) questions the application of this model and its variants such as SECI Model (Socialization, Externalization, Combination, Internalization) proposed by (Nonaka & Konno, 1998) in rapidly changing environment characterized by technology discontinuity such as software development.

Network model is based on socialization of knowledge and relationships of actors; the model highlights the role of social patterns between individuals and interest groups in knowledge creation, sharing and transfer. It has the advantage of focusing on external sources of knowledge through interest and practice networks. It inoculates the collaborative aspect of creating knowledge and sharing which is a key factor in software development especially in geographically dispersed teams (Hemetsberger & Reinhardt, 2003). Community of practice (CoP) model -: the term CoP was coined by Jean Lave and Etienne Wenger, who described it as "groups of people informally bound together by shared expertise and passion for a joint enterprise"(Wenger & Snyder, 2000) .The work of (Sharratt & Usoro, 2003) differentiated CoP from usual work teams and organizational functional units in that they are self organizing systems and their existence is guided by membership perceived gains. These communities are not constrained by time and space and can span beyond organizational boundaries.

Swan & Newell, (2000) contend that trust based rules of engagement are a critical factor to the success of this model. This model provides a good background for KM initiatives in software engineering especially open source development.

Quantum Model is based on recent advances in quantum computing, the assumes that application of quantum computing to the constituents of knowledge will lead to high level complexity and improved rationality in decision making as actors in given scenarios in a the context of application. This model is not appropriate for use in low resourced communities.

*Knowledge Management in Software Development*
There are two scenarios on human generated uncertainty in the software development (Dekhtyar, Hayes, & Goldsmith, 2007). They include:

Uncertainty on the process which includes issues like : How long will it take?, what is the most efficient development methodology?, the choice of language and environment

Uncertainty on the product which includes issues like: How much of security features is required (the tradeoff between usability and security).There is increasing use of Global Software Development (GSD) teams inform of globally distributed subsidiaries of the same organization, outsourced companies, open source communities or collaborating virtual companies which are distributed globally working on complex software projects (Avram, 2007; Hemetsberger & Reinhardt, 2003). Integration of KM in the Software development environment context can be used to improve on the quality of the product (process output), the process quality itself and reduce on uncertainty associated with software development. Software development regardless of the nature and the level of uncertainty is collaborative and requires intensive human decision especially when adaptive development methodologies are used (Dekhtyar et al., 2007). Throughout the software development lifecycle, collaboration is among actors with differing expertise.

We formulate three scenarios that illustrate the different levels of differing expertise that result from level of experience gained through practice and/or Interest: *Scenario 1:* Where the customer is a research hospital in need of Hospital Management software, the customer may understand his domain well i.e. medicine and hospital management but may have zero knowledge on software domain. The customer may even have the knowledge on the application of softwares in his domain but not the development of softwares. *Scenario 2*: In an open source community developing antivirus software may have medical doctors with interest (Community of Interest) in the study of computer viruses and other malicious software, they may have Zero knowledge on the software development but their domain expertise on viruses is necessary for the success of the project. *Scenario 3:* In a Global Software Development (GSD) all stages in the software development lifecycle are carried in culturally diverse environments, the software engineers may have differing experiences in the practice of software engineering.

*The Hybrid Framework for KM in Software Engineering*
To cater for these unique circumstances as illustrated through the three scenarios given above, for example, the presence of both communities of interest and communities of practice. We therefore, propose a hybrid framework for KM process that blends philosophical, cognitive, CoP and network models and can be effective in software development environment.

The Figure below shows the proposed hybrid framework. The stages of knowledge management in software development within the framework are: Knowledge Creation ; Validation and Audit; Transfer; Consolidation of best practices; Documentation

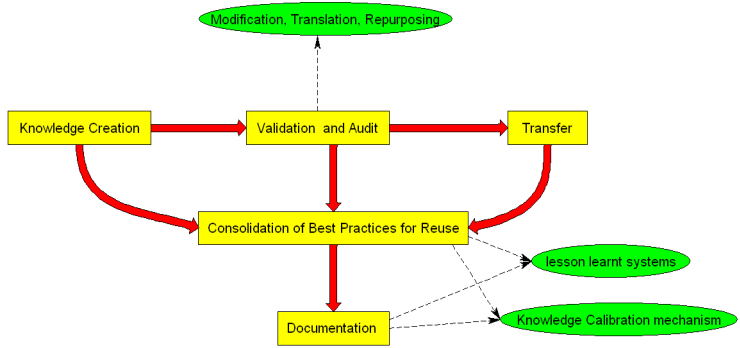

In the next sub-sections, we discuss the granules that make up each of the stages and highlight its building blocks of achieving a comprehensive KM process in the software development and its probable implementation in East Africa.

*Knowledge Creation*

Knowledge creation is the most important area of focus within knowledge management since this stage inputs have far reaching effects on the preceding stages of the KM framework (Wickramasinghe, 2006).

Knowledge can be created by people and/or technologies or be embedded in processes as shown in the Figure below KM Triad adapted from (Wickramasinghe, 2006).

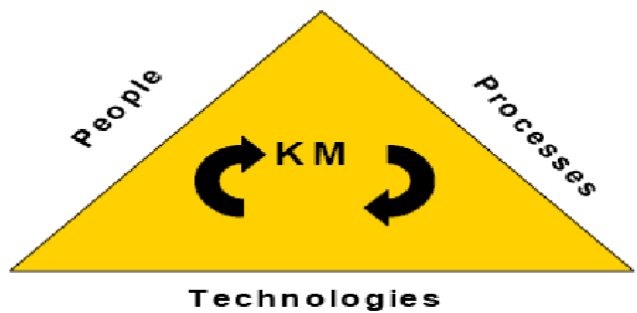

In GSD, knowledge creation involves virtual teams each made up of the three aspects people, technologies and processes who can either be localized or distibuted. The knowledge creation process in group environment has be dealt with in depth by (Drach-Zahavy & Somen, 2001; Gibson, 2001; Mitchell & Nicholas, 2004, 2006) (Mitchell & Nicholas, 2006).Their studies identifies four group knowledge creating processes . The first process is accumulation of knowledge on individuals originating from their functional areas or practice. The authors identify only the communities of practice as the probable affiliation of these individuals. In software development, the communities of interest have significant role in building shared knowledge base. These individuals who have interest in the software environment are not bounded by practice but context of interest. In our proposed framework both communities are considered.

The second Process is Interaction; this involves team sharing the accumulated individualistic knowledge to create new individual and team knowledge. Since Knowledge is context dependant, members of localized or virtual teams are influenced by their interest to participate in this process. In software development and virtual teams this process involves collaborative technologies such as groupware. Analysis is the third process of group knowledge creation; in this process group members explore their experiences in comparison with other team members. Group analogical reasoning in software engineering have numerous applications across the development lifecycle such as requirements engineering where customers view the system largely from the usability perspective (usability requirements) are able to share with developers.

The fourth process is integration and creation which encapsulates consensus building on the experiences and analogies. It can be facilitated through story telling. To sustain the group knowledge creation and the preceding stages of the hybrid framework, two major continuous activities are initiated:Laying foundation for building knowledge taxonomies.(Whittaker & Breininger, 2008) provides a detailed approach on building taxonomies;Community Social Network Analysis (SNA) for measuring centrality and expert localization. The study by (Dekker & Hendriks, 2006) lay much emphasizes on Social Network Analysis.

*Knowledge Validation and Audit*
Various roles and tools are emergent in the process of building taxonomies and expert localization. Using an organizational perspective of KM, there are various roles and tools that may emerge in the organization to carry out key steps in KM validation and audit (Modification, translation and Repurposing (Rao, 2005). (See Figure below: Madanmohan Rao on Roles online Communities)

| Roles and Tools for Online Communities | | |
|---|---|---|
| Knowledge roles | Activities | Tools |
| Knowledge consumer | Search, browse, access, apply, learn | Portal, search engine, workflow |
| Knowledge creator | Publish, improve, classify, discuss | Content management, authoring, taxonomy, online CoPs |
| Knowledge editor | Interviewing experts, storytelling, content management | Content management systems, taxonomy |
| Knowledge expert | Validate, certify, legitimize | Online CoPs, ranking/rating tools, best practice repository |
| Knowledge broker | Locate experts/knowledge, identify gaps, organize, filter, coordinate CoPs | Enterprise portal, audit tools, online forums, organizational knowledge maps |
| Knowledge leader | Shape KM agenda, align with business objectives | Intellectual capital navigators, industry knowledge maps |

There are other authors who discuss an organizational and a community view of knowledge management (Bourhis, Dubé, & Jacob, 2005; Fontaine, 2001). They illustrate various roles in knowledge validation and audit, highlighting the role of leadership as a critical pillar in Online Community Knowledge Validation and audit see Figure below; Typology of community roles .adapted from (Bourhis et al., 2005; Fontaine, 2001)

| | Role | Description |
|---|---|---|
| | Community Members | Take active ownership in the community by participating in its events and activities and driving the level of commitment and growth of the community. |
| Leadership Roles | Community Leaders | Provide the overall guidance and management needed to build and maintain the community, its relevance and strategic importance un the organization and level of visibility. |
| | Sponsors | Nurture and provide top-level recognition for the community while ensuring its exposure, support, and strategic importance in the organization. |
| Knowledge Intermediary Roles | Facilitators | Network and connect community members by encouraging participation, facilitating, and seeding discussions and keeping events and community activities engaging and vibrant. |
| | Content Coordinators | Serve as the ultimate source of explicit knowledge by searching, retrieving, transferring and responding to direct requests for the community's knowledge and content. |
| | Journalists | Responsible for identifying, capturing, and editing relevant knowledge, best practices, new approaches and lessons learned into documents, presentations and reports. |
| Community Support Roles | Mentors | Act as community elders, who take a personal stake in helping new members navigate the community, its norms and policies and their place in the organization. |
| | Admin/Events Coordinators | Coordinate, organize and plan community events or activities. |
| | Technologists | Oversee and maintain the community's collaborative technology and help members navigate its terrain. |
| Knowledge Domain Roles | Subject Matter Experts | Keepers of the community's knowledge domain or practice who serve as centers of specialized tacit knowledge for the community and its members. |
| | Core Team Members | Looked upon for guidance and leadership before or after a leader emerges or is selected; guidance includes developing the community's mission and purpose. |

Techniques such as members' contribution valuation through community assessment, rating/ ranking/ scoring can be used. This is a common practice in existing online user communities.

*Knowledge Transfer*

Effective transfer of Knowledge between knowledge workers is one of the key challenges in KM (Alavi & Leidner, 2001; Joshi, Saonee.Sarker, & Sarker, 2004) . (Jacobson, 2006) defines knowledge transfer as: *"An exchange of knowledge in which the focus is on structural capital (knowledge that has been built into processes, products, or services) and on the transformation of individual knowledge to group knowledge or organizational knowledge"*

In tandem with our proposed framework , (Jacobson, 2006) definition reinforces the view that there must exist a normative structure that provides for knowledge flow in the intended community. (King, 2006) gives an organizational view in the transfer process and argues that knowledge transfer is effective if the sender and receiver are in homogeneous contexts. Self motivation of knowledge workers, awareness and acceptance of the transfer goals are some of factors positively influence knowledge transfer.

*Documentation and consolidation of best practices.*

These stages cater for the future use of assimilated knowledge. In software development, knowledge reuse can be seen as an extension of components re-use that reduce on new software development costs. Lesson learnt systems can be used to catalogue the experiences gained through all the stages of knowledge management for knowledge reuse. Knowledge calibration is the correspondence between accuracy of knowledge and confidence of knowledge as a reliable base where knowledge workers can confidently base their decision (Goldsmith & Pillai, 2006). A detailed literature on knowledge calibration is available in (Goldsmith & Pillai, 2006).

## Conclusions and Areas of Further Research

The paper begins with an introduction to KM and a brief review of KM initiatives. Then an exploratory literature study on existence of KM initiative is carried out on software development in East Africa which focuses beyond the organizational view. The study establishes there is no formal study or open initiative for KM in software development in the region. Based on these findings, we explored the existing major models of KM on their viability and application in software development. Following an assessment of individual models, we propose a generic framework that blends the four major models. The proposed framework is intended to be the starting point for KM initiative among stakeholders in the software industry in East Africa. The authors discuss the various stages of this framework and their output. We confer that the use of this framework can help the software developers in East Africa create, use and share valuable experiences that will give them competitive advantage in the global market. What remains to be seen is how this framework can be incorporated to formalize the KM initiative among the software development community. We suggest research be undertaken for determining: Comparative effectiveness of the proposed framework in the development of local mobile content; The benchmarking standards for best practices. An empirical study can be conducted; The viability of developing and using open source technologies that extends beyond cultural barriers such as use of natural language processing facilities at various stages of this framework.

**Brief Bio-data about the authors**:

Karanja Evanson holds an Msc in computer science (Makerere University) with a bias in machine learning and security, among other qualifications. He is an ICT enthusiast, a Lecturer in ICT and an entrepreneur. His research interests includes, programming languages, E- government, knowledge management and Information security.

Lawrence Xavier Thuku : Thuku has Master of Science in Information Technology (MSc. IT) from Strathmore University and various ICT industry Certifications .He is Lecturer at the Institute of advanced Technology and a consultant . His research interests includes Workflow Management systems , Decision Support systems and knowledge management.

John P. Kangethe Karanja holds a MBA (Finance) from University of Nairobi , CPA and CPS among other qualifications. He has taught Finance and accounting courses at various Universities in Kenya.